\documentclass[fleqn,twoside]{article}
\usepackage{espcrc2,amssymb}

\usepackage{graphicx}
\usepackage[figuresright]{rotating}

\newcommand{\AmS}{{\protect\the\textfont2
  A\kern-.1667em\lower.5ex\hbox{M}\kern-.125emS}}

\title{High Energy Neutrino Astrophysics}

\author{Teresa Montaruli\address{Universit\`a di Bari and INFN, 
        Via Amendola, 173,  
        I-70126, Bari, Italy, montaruli@ba.infn.it}}
       
\begin{document}

\begin{abstract}
Astroparticle physics is a field rich of perspectives 
in investigating the ultra-high energy region which will never be accessible
to man made machines. Many efforts are underway in order to detect 
the first cosmic neutrinos, messengers of the unobserved universe. 
The motivations of this field of research, the current and next future
experimental status are reviewed. 
\end{abstract}

\maketitle

\section{Introduction}

The observation of neutrinos is an experimental challenge
due to their weak interactions that makes them elusive particles.
%Nevertheless nowadays neutrino detection is providing numerous results
%as demonstrated by the 2002 Nobel Prize recognized to 
%Profs. R. Davis and M. Koshiba for inaugurating the field of 
%neutrino astrophysics. 
Though they are difficult
to detect, they are excellent probes to
observe the most distant sources and their interior. 
%because they are weakly absorbed.
Until now only MeV extra-terrestrial
neutrinos have been observed from the Sun and SN1987A.
Nevertheless there is growing activity to build detectors 
for high energy ($\gtrsim 1$~TeV) neutrinos of astrophysical origin: 
their detection would allow
the observation of the horizon not accessible to the well-established
$\gamma$-astronomy.

In Sec.~\ref{sec:motiv} the scientific objectives of $\nu$
astrophysics and the connection to the observation
of Ultra-High Energy Cosmic Rays (UHECRs) are introduced. The complementarity
between $\gamma$ and $\nu$-astrophysics is pointed out and
typical models for $\nu$ production are considered.
The working principle of Neutrino Telescopes (NTs) exploiting the
Cherenkov light detection in the South Pole ice or in sea/lake depths is 
summarized in Sec.~\ref{sec:nt}. Detector performances, event topologies
and background rejection methods are considered.
%Expected rates are modest from typical models, justifying the need for
%large detectors. 
In Sec.~\ref{sec:exp} the current status of Cherenkov 
NTs is reviewed, while in Sec.~\ref{sec:other} other
techniques are considered. 

\section{Scientific Motivations of $\nu$ Astrophysics
%: the cosmic ray and $\gamma$-ray connection
}
\label{sec:motiv}

Cosmic rays have been observed in huge Extensive Air Shower (EAS)
arrays up to energies in excess of $10^{20}$~eV, exceeding those of
any foreseen accelerator machine. 
The spectrum is impressively regular: 2 power laws with
a mass-dependent break at $\sim 3000$~TeV (the 'knee')
beyond which the spectrum becomes softer. It is commonly believed that
up to the knee galactic supernovae (SNs) can be the sources of CRs,
since they have the proper power to
accelerate CRs and to balance the energy density of CRs confined
in the Galaxy. Moreover diffusive shock acceleration naturally leads
to power law spectra. Most experiments agree that above the knee the
composition becomes heavier since the proton gyro-radius in the
galactic magnetic fields is smaller than that of Fe nuclei.
Another feature at $\sim 10^{19}$~eV
(the 'ankle') indicates another possible spectral slope change
not yet well established due to the small statistics ($\sim 1$ 
particle/km$^2$/yr). At these energies the Larmour radius of
protons  is comparable to the Galaxy size and the ankle
could be due to the onset of an extra-galactic component.
Most of the EAS observations 
agree on a proton dominated composition. Nevertheless it is still
not possible to establish the fall-off of the CR spectrum above
$10^{19.5}$~eV (GZK cut-off),
due to p interactions with the Cosmic Microwave Background Radiation (CMBR),
since statistics is small and systematic errors large. 

Since $\gamma$-rays are reprocessed in sources and absorbed by extra-galactic
backgrounds (IR, CMB and Radio photons) through pair production, at few 
hundreds of TeV they do not survive the journey from the Galactic Center.
Protons cannot access regions at distances 
$\gtrsim 50$~Mpc at energies E$>50$ EeV at which they are deflected by 
$<1^{\circ}$ by magnetic fields. Neutrons of EeV energies have
decay lengths of the order of 10 kpc. Hence, there is no doubt that
neutrinos, and possibly gravitational waves, propagating
almost undisturbed by encountered matter and magnetic fields, 
represent the 'probes' of the
Universe with unique discovery potentials. They could allow us
to answer the question: {\it 'Which are the sources of the
highest energy cosmic rays?'}. Review papers on $\nu$ astrophysics are
in \cite{GHS}.

Models for high energy CR production and acceleration are divided into 2 
classes: top-down models, that imply super-massive relic decays,
%can provide energies up to the grand-unified mass of $10^{24}$~eV,
and bottom-up models. In bottom-up scenarios protons or nuclei are accelerated
and interact with matter or photons surrounding
the accelerator, hence producing mesons.
Neutral pions decay into $\gamma$s and from charged pion decay (if all
$\mu$'s decay too) the $\nu$ species are produced in a flavor
ratio of $\nu_{e}:\nu_{\mu}:\nu_{\tau}=1:2:0$, since $\nu_{\tau}$ production
at source from charmed mesons is negligible. Moreover after propagation
through cosmological distances, these ratios are altered by $\nu_{\mu} 
\rightarrow \nu_{\tau}$ oscillations into $1:1:1$ (even though other scenarios
involving $\nu$ decay are possible). 
%and could be tested in $km^3$ NTs).
From these reactions  
the strict relation-ship between gamma and 
neutrino astrophysics emerges. It is reasonable to assume that at the sources 
photon and
neutrino spectra have similar shape and normalization (even if absorption 
effects can be not negligible), typically
$\frac{dN}{dE} \propto E^{-(2 \div 2.5)}$, as expected from $1^{st}$ order
Fermi acceleration mechanism.
Irrespective of details of acceleration mechanisms, 
the maximum energy of CRs accelerated in a site of dimensions $R$
in the presence of magnetic fields B can be calculated
considering that the Larmour radius of particles 
cannot exceed $R$. It results in $E_{\max} \sim
\beta Z \left( \frac{B}{1 \mu {\rm G}} \right) \left( \frac{R}{1 {\rm kpc}}
\right) 10^{18}$ eV, where $\beta$ is the velocity of a shock wave 
in units of velocity of light or 
the efficiency of the acceleration mechanism. 
Other $\nu$ production mechanisms are photopion production due to
UHECR interactions with the CMBR 
(`GZK $\nu$'s') \cite{GZKnu} 
and Z decays due to UHE $\nu$'s interacting on 
cosmic background $\nu$'s \cite{Weiler}.

Examples of sources accelerating CRs up to energies 
of $10^{19}$ eV are extra-galactic sources, such as
inflows onto super-massive black holes at centers of 
Active Galactic Nuclei (AGNs) and Gamma-Ray bursters (GRBs). 
If the sources of the observed highest energy CRs are optically
thin to proton photo-meson and proton-nucleon interactions,
so that protons escape,  
an upper limit to the $\nu$ flux (hereafter W\&B limit)
can be calculated \cite{WB}. This flux 
corresponds to $\sim 200$ events/km$^2$/yr.
It should be considered that the W\&B flux is derived normalizing the
observed proton flux at 10 EeV and extrapolating it at lower
energies using an $E^{-2}$ source spectrum. It can be evaded 
considering other spectral
dependences; moreover 
magnetic field effects and uncertainties in photohadronic interactions
can reduce the number of protons able to escape, hence
affecting the limit \cite{MPR}.
Galactic sources are not subject to these limits due to their
proximity. Possible sources of HE $\nu$'s
are micro-quasars, exploding stars and consequent neutron stars
in SNRs, magnetars and the rates vary from fractions of
events up to hundreds in km$^2$ detectors.

There is not a convincing evidence nowadays on the existence
of hadronic acceleration in sources. The TeV gamma sky is of
great interest for $\nu$ astrophysics since if a
strong indication emerges that $\gamma$s are not produced by electromagnetic
processes but in $\pi^0$ decay, HE $\nu$'s should
be produced as well by charged $\pi$s. 
%The imaging atmospheric Cherenkov (IAC)
%technique is an effective tool in detecting $\gamma$ emissions from
%souces at energies $\gtrsim 300$ GeV and 
Current TeV catalogues include about 20 sources detected
by imaging atmospheric Cherenkov (IAC) detectors, 
mainly extra-galactic BL Lacs and galactic SN remnants (SNR). 
%More recently 3 new classes of objects were found, i.e. starburst galaxies 
%(NGC 253), radio galaxies (M87) and OB associations (Cyg OB2). 
CANGAROO IAC experiment 
has detected at $\sim 7\sigma$ level TeV emissions from
the Galactic Center \cite{ICRC2003gamma}, 
and it is expected to be confirmed by HESS soon. 
The claim by CANGAROO \cite{CANGAROO} on the better 
compatibility of TeV $\gamma$-ray measurements from the SNR 
RX J1713.7-3946 with a $\pi^0$ decay hypothesis
compared to electromagnetic mechanisms (bremsstrahlung and inverse Compton)
is still controversial 
%and could be in disagreement with EGRET upper limit 
\cite{Pohl}.
There is room for a $\pi^0$ decay contribution in HE tails 
of multi-wavelength spectra 
of some TeV sources, such as the Crab, PSR 1706-44, 
Cas A, plerions and shell-type SNRs.
\section{Cherenkov Neutrino Telescopes
%: detection principle and expected rates
}
\label{sec:nt}

Neutrinos can be detected through their charged current (CC) interactions
in 3-D arrays of optical modules (OMs), 
pressure resistant glass spheres containing phototubes (PMTs),
located in polar ice or sea/lake water.
OMs distances are optimized considering light transmission properties 
and construction constraints. 
The time and position of PMTs hit by Cherenkov light emitted by 
relativistic particles allow the reconstruction of tracks.
Charge amplitudes are used to measure $\mu$ and shower energies.
The selection of detector sites is determined by transmission light
properties, environmental
backgrounds, stability of media properties in the implemented region, 
mechanical, construction and infrastructure constraints. 
In Sec.~\ref{sec:exp} merits and drawbacks of ice and
water properties are discussed.

For various reasons this technique improves with energy.
$\nu$ cross-sections and $\mu$ range increase with energy,
and hence the effective target mass.
The amount of light also increases hence reconstruction
of $\mu$ tracks and of cascades induced by $\nu_{e,\tau}$ and NC can improve.
%Cascades are point-like given the typical granularity of sensors,
%the sensitive volume is the implemented region enlarged by a few 
%light attenuation lengths.
Moreover the signal to noise ratio (S/N) improves with energy, since
the atmospheric $\mu$ and $\nu$ fluxes are steeper ($\sim E^{-3.6}$ for 
$E \gtrsim 100$ GeV) than fluxes expected from sources ($\sim E^{-2}$).

In order to reduce the $\mu$ flux by orders of magnitude,
detectors are located below kms of matter, underwater or underice.
%to make use of natural Cherenkov media.
The rejection of atmospheric $\mu$'s is achieved looking at events from
the lower hemisphere, induced by $\nu$'s crossing
the Earth, or using HE cuts in case also events from above
should be recovered from the background. 
In fact, at $\sim 40$ TeV the $\nu_{\mu}$ 
interaction length equals the Earth diameter and at $E \gtrsim 1$ PeV
$\nu_{\mu}$ absorption is severe. On the other hand, the Earth is transparent
to $\nu_{\tau}$ thanks to $\tau$ decay producing another $\nu_{\tau}$. 
Hence NTs need to have good shower, track reconstruction
capabilities and energy resolution.

The atmospheric $\nu$ background rejection in astrophysical $\nu$ searches 
depends on strategies.
For point-like sources, statistically significant 
clusters of events with respect to the atmospheric $\nu$ distributions
are looked for.  Methods are based on angular cuts optimized in order
to have the best S/N ratios
in case of binned methods or on the measurement of the
energy dependent point-spread function for unbinned methods. 
As a matter of fact, the distribution of events around the
point source is different from the background. 
In case of time-dependent emissions, such as GRBs, further time 
requirements strongly reduce the backgrounds.
It is clear that a relevant parameter for
point-like source searches is the angular resolution.
% $\Delta\theta$ since
%the sensitivity is proportional to $\sqrt{AT}/\Delta\theta$, where
%$A$ is the effective area and T the detection time.
For galactic sources, for which typically $\nu$ emissions do not
exceed energies much larger than 100 TeV, a good angular resolution provides
a substantial mean to reject the backgrounds. 
On the other hand, for diffuse extra-galactic sources 
the S/N ratio is optimized using minimum 
energy cuts and exploiting the different energy dependence 
of signal and noise spectra. 
Both $\mu$ tracks or cascades are used for diffuse 
flux searches. Typically, the direction of $\nu$ parents of cascades  
is detected with worse resolution than for
$\mu$'s (typically $\lesssim 30^{\circ}$
above 10 TeV), while the energy resolution is competitive (for $\mu$'s
$\sim 30-40\%$
in logE, for showers $\sim 20\%$ in E above 10 TeV).

The effective area for $\nu$'s, that is the sensitive area
'seen' by $\nu$'s producing detectable $\mu$'s
when entering the Earth, is a useful parameter
to determine event rates and to be compared between 
experiments. In fact, the event rate for a $\nu$ model predicting
a spectrum $\frac{d\Phi}{dE_{\nu}d\Omega_{\nu}}$
is given by $N_{\mu} = \int\int dE_{\nu}d\Omega_{\nu} 
A^{eff}_{\nu}(E_{\nu}, \Omega_{\nu})
\frac{d\Phi}{dE_{\nu}d\Omega_{\nu}}$. It depends on
track reconstruction
quality cuts and selection criteria for 
background rejection, on the $\mu$ range, $\nu$ cross
section and on the $\nu$ absorption in the Earth. 
%Another useful parameter, is the effective area for $\mu$'s,
%defined as the ratio of selected event rate
%to the incident flux on the detector, is the main parameter determining 
%the $\mu$ induced event rate. 
%It depends on $\mu$ energy and on selection criteria on track reconstruction
%quality and background rejection. 
Being strongly energy dependent,
%detectors respond in different regions to $\nu$'s with different spectra:
the mean energy of atmospheric $\nu$'s producing detectable events  
is $\sim 100$ GeV, while for an $E^{-2}$ spectrum it is $\sim 10$ TeV. 
%Apart from atmospheric $\mu$'s and $\nu$'s 
%other backgrounds are of environmental nature (See Sec.~\ref{sec:exp}). 

\section{Experimental Status and Results}
\label{sec:exp}

The NT currently taking data are AMANDA at the South Pole \cite{AMANDA}
and NT200 (192 OMs on 8 strings) in Lake Baikal (Siberia) \cite{Baikal} 
at 1.1. km depth.
Baikal effective
area for $\mu$'s is 2000 m$^2$ at 1 TeV and the sensitivity to cascades
is competitive to AMANDA.
By implementing 3 additional strings
carrying only 6 couples of OMs vertically spaced by 70 m,
at a distance of 100 m from the detector centre (of diameter of 43 m)
the sensitivity to cascades will improve by a factor of 4. Baikal
was the first underwater telescope to reconstruct atmospheric $\nu$'s
in 1996.

\begin{figure}[htb]
  \begin{center}
    \includegraphics[height=15.pc,width=15.pc]{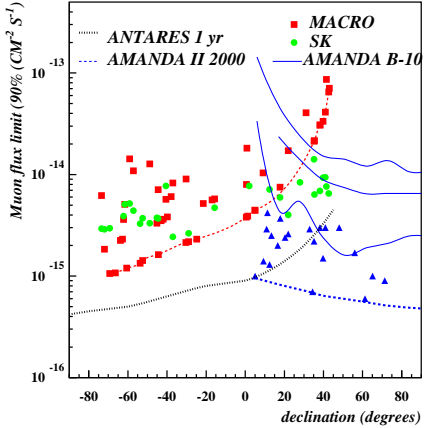} 
  \end{center}
\vskip -1.3cm
  \caption{90\% c.l. upper limits on $\mu$ fluxes induced 
by $\nu$'s with $E^{-2}$ spectrum vs source
declination for SK (green circles) \protect\cite{SK}, 
MACRO (red squares) (updated 
with respect to \protect\cite{MACRO}), AMANDA-B10 
%(3 blue solid lines:
%the upper and lower lines indicate the excursion of limits in
%different R.A. bins for a given declination band) 
\protect\cite{AMANDAB10}, 
AMANDA-II 2000 data (triangles),
% \protect\cite{AMANDA2},
ANTARES sensitivity (dotted black line) after 1 yr \protect\cite{ANTARES}.
It has not been possible to apply a correction due to different
$\mu$ average energy thresholds (SK $\sim 3$~GeV, MACRO
$\sim 1.5$ GeV, AMANDA $\sim 50$~GeV). Nevertheless, the 
maximum of the response curves of these
detectors for an $E^{-2}$ flux is at $E_{\mu} \sim 10$~TeV, hence 
events contributing between 1-50 GeV should not
make a large correction to these limits. 
}
\vskip -0.5 cm
\label{fig:point}
\end{figure}

AMANDA is running now in the AMANDA-II configuration with 677 OMs with
8-inch PMTs on 19 strings
implemented between 1.5-2 km deep in the ice.
The angular resolution for $\mu$ tracks has improved from 
$\sim 4^{\circ}$ for the previous configuration AMANDA-B10 \cite{AMANDAB10}
(302 OMs on 10 strings) to $2-2.5^{\circ}$. 
The effective area for $\mu$'s, that
has largely improved in the horizontal direction, is 
$\sim 0.02-0.04$ km$^2$ for an $E^{-2}$ $\nu$ flux depending on the source 
declination. 
From the two `calibration' test beams of atmospheric $\mu$'s and $\nu$'s 
a systematic error of 25\% on detector acceptance
is derived mainly due to OM sensitivity and
ice optical property knowledge.  
The upper limits for $\mu$ fluxes for point sources
calculated using 699 upward events 
selected in 197 d during the year 2000, 
are shown in Fig.~\ref{fig:point}. The most significant excess, observed 
around 21.1h R.A. and $68^{\circ}$ declination, 
is of 8 events and the expected background is 2.1. The probability 
to observe such an excess as a random fluctuation 
of the background is 51\%.
Limits from other experiments of smaller area 
($\sim 1000$ m$^2$), Super-Kamiokande (SK)\cite{SK} and MACRO \cite{MACRO}
in the upper hemisphere, are reported.
It is interesting to notice that the MACRO scintillator+tracking detector, 
with angular resolution $\lesssim 1^{\circ}$, 
in a sample of 1388 upward-going $\mu$'s
has found an excess in the Circinus Constellation region.
Ten events have been detected inside a $3^{\circ}$ half-width cone
(including $90\%$ of an $E^{-2}$ signal) around the plerion PSR B1509-58
and 2 are expected from atmospheric $\nu$'s. This source was also
detected by CANGAROO in 1997 above 1.9 TeV with
$4\sigma$ significance, but not confirmed
by 1996 and 1998 data with 2.5 TeV threshold
\cite{PSRB}. Even though PSR B1509-58 is of interest as possible $\nu$
emitter, the significance of MACRO result is negligible 
when all the 1388 directions of the measured events are looked at. 
Moreover, it is
expected that $E^{-(2 \div 2.5)}$ signals should produce at least 4-7 events 
in $1.5^{\circ}$ around the source while only 1 is detected and the expected
background is 0.5.  
From the same sources SK observes the largest number of events between the
selected catalogue looked at, but the data are still
compatible with background fluctuations
(9 events to be compared to a background of 5.4).
The sensitivity of ANTARES 
(expected angular resolution of $\sim 0.2^{\circ}$ for $E_{\nu}>10$ TeV) 
is also shown for 1 yr of data taking \cite{ANTARES}. 

\begin{figure}[htb]
\begin{center}
    \includegraphics[height=15.pc,width=15.pc]{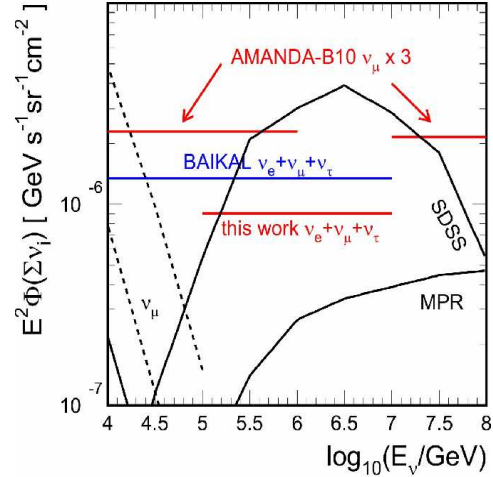} 
  \end{center}
	\vskip -1.3cm
  \caption{90\% c.l. upper limits on diffuse $E^{-2}$ $\nu$ flux
for cascades of all flavors. The AMANDA-II (197 d) 
\protect\cite{AMANDAcascades} and Baikal (268 d)
\protect\cite{Baikal} results are presented compared to atmospheric 
$\nu$ fluxes,
an AGN model (SDSS) and an upper limit on $\nu$ fluxes \cite{MPR}). 
For comparison the AMANDA B-10 $\nu_{\mu}$
limits are shown multiplied by a factor of 3 (the underlying hypothesis
is that at source the proportion between flavors is $\nu_e : \nu_{\mu} :
\nu_{\tau} = 1:2:0$, while at Earth, due to oscillations,
it is $1:1:1$).}
\label{fig:cascades}
\end{figure}

Upper limits for $E^{-2}$ 
diffuse fluxes of $\nu_{\mu}+\bar{\nu}_{\mu}$ are summarized
in Tab.~\ref{tab:diffuse}.  
When specific spectral shapes of models are considered,
it results that some $\nu$ models from AGNs/blazars and quasars 
are excluded by the AMANDA B-10 limit, 
while this is still higher than the W\&B limit of $4.5 \cdot
10^{-8}$ GeV cm$^{-2}$
s$^{-1}$ sr$^{-1}$.
Upper limits on diffuse fluxes of cascades induced by
all flavor $\nu$'s 
are shown in Fig.~\ref{fig:cascades}.
\begin{table*}[htb]
\caption{90\% upper limits on $\mu$ fluxes 
above $E_{\mu \, min}$ and to diffuse $\nu_{\mu}+\bar{\nu}_{\mu}$ $E^{-2}$
fluxes in the given energy intervals. For ANTARES (error due
to prompt $\nu$ prediction uncertainties) and IceCube the 
expected sensitivities are given. $\mu$ track reconstruction is required
(except for the UHE limit by AMANDA B-10 for showering 
$\mu$'s obtained using neural network methods) and energy cuts are
used to reject atmospheric backgrounds.}
\label{tab:diffuse}
\newcommand{\m}{\hphantom{$-$}}
\newcommand{\cc}[1]{\multicolumn{1}{c}{#1}}
\renewcommand{\tabcolsep}{.05pc} % enlarge column spacing
\renewcommand{\arraystretch}{1.2} % enlarge line spacing
\begin{tabular}{@{}lllll}
\hline
Experiment           & Run time & Energy & $\Phi_{\mu}$ & $\Phi_{\nu}$ \\
Reference	     &  (yrs) &range & cm$^{-2}$ s$^{-1}$ sr$^{-1}$ &
GeV cm$^{-2}$ s$^{-1}$ sr$^{-1}$ \\
\hline
AMANDAB-10\protect\cite{AMANDAdiffuse}  & \m0.36 & $E_{\nu}\sim
6-10^3$~TeV & & \m$8.4 \cdot 10^{-7}$  \\
AMANDAB-10 \protect\cite{AMANDAUHE}  & \m0.36 & $E_{\nu}\sim
2.5-5.6\cdot 10^3$~PeV & & \m$7.2 \cdot 10^{-7}$  \\
MACRO \protect\cite{MACROdiffuse}  & \m5.8 & $E_{\nu}(E_{\mu \, min})$
$\sim 10^4-10^6$ (1.5)~GeV& 
\m$1.7 \pm 0.2 \cdot 10^{-14}$  & \m$4.1 \pm 0.4 \cdot 10^{-6}$\\
Soudan2 \protect\cite{SoudanAGN} & \m6.34 
& $E_{\mu \, min}\sim 5-20-100$~TeV& $2.2-1.5-1.4\cdot 10^{-14}$ &\\
Frejus \protect\cite{Frejus} & \m3.27 &\m$E_{\nu} \sim 2.6$ TeV&
& $4.7 \cdot 10^{-6}$ \\
ANTARES \protect\cite{ANTARES} &\m3 &\m$E_{\mu \, min} = 125$~TeV
& & $3.9 \pm 0.7 \cdot 10^{-8}$\\ 
IceCube \protect\cite{IceCube} & \m3& \m$E_{\nu}\gtrsim 100$ TeV & 
&$4.2 \cdot 10^{-9}$\\
\hline
\end{tabular}\\[2pt]
\end{table*}

ANTARES \cite{ANTARES} will be located in front of Toulon, South France, 
40 km off-shore at a depth of 2400 m. It will consist of 12 strings
carrying 75 OMs each, containing 10-inch PMTs. 
The electro-optical cable (EOC) already
connects since Dec. 2002 the shore station to the junction box, that
distributes data and power to strings. Submarine connections were
successful in Mar. 2003 when a prototype (1/5 of a string with 15 OMs),
was deployed together with an instrumentation string for 
environmental parameter measurements. The detector is expected to be
completed by the end of 2006.

NESTOR \cite{NESTOR} is aiming at the construction of a NT 
close to Pylos (Greece) at $\sim 4$ km depth. Proposed towers are made of
12 hexagonal floors of 32 m diameter spaced by 30 m each carrying 6
upward-looking and 6 down-ward looking OMs with 15-inch PMTs. 
The effective area for $E_{\mu}> 10$~TeV 
is $\sim 0.02$~km$^2$. In Mar. 2003 a prototype 12 m diameter
floor was deployed and PMT data were transmitted to shore trough a 35-km
EOC.

In the next future the community is aiming at the construction
of detectors at the scale of km$^3$.   
The IceCube project \cite{IceCube} is already funded 
and detector construction will start in the Austral summer of 2004-5 
and will continue for about 6 years.
It will consist of 4800 DOMs (digital OMs) 
on 80 strings each with 60 10-inch PMTs
vertically spaced by 17 m extending from 2.4 km up to 1.4 km depths.
The strings are at vertices of equilateral triangles with 125 m long side.
Close to each string hole there will be 2 iced water tanks seen by 2 DOMs,
forming the IceTop array for CR composition measurements and 
absolute pointing determination. 
The $\mu$ declared effective area after selection requirements
is $>1$ km$^2$ above 10~TeV, the
angular resolution is expected to be $<1^{\circ}$ at high energies,
the energy resolution $\sim 30\%$ in $\log E_{\mu}$ and $\sim 20\%$ 
in E for cascades. 
%, critical to reject atmospheric $\mu$ and $\nu$ backgrounds and recover 
%events from the upper hemisphere at energies in excess of 1 PeV, when
%most of the upward $\nu_{\mu}$'s are absorbed by the Earth, 

In order to cover the entire sky, particularly the region of the
Galactic Center, a $km^3$ detector is envisaged also in the Mediterranean.
Being located in sea water, it will provide complementarity respect to 
IceCube for what concerns media properties and environmental backgrounds. 
Media transmission properties are characterized by the attenuation length,
the sum of the absorption and scattering lengths. 
Absorption reduces signal amplitudes, hence it 
mainly affects the choice of the distance between OMs. Scattering
length affects the direction of light propagation and the
arrival time of photons on OMs, hence the angular resolution. A useful
parameter is the effective scattering length which takes into account
the angular distribution of scattered photons. It was found by the AMANDA
experiment that ice properties depend on depth: the presence of
air bubbles in the ice is much reduced below depths of 1 km, and
at the depths in which AMANDA is currently located (1.5-2 km) the
scattering is  
$\lambda^{eff}\sim 25$~m, even though the ice property dependence
on depth and the effect of bubbles that form around OMs after drilling,
is still one of the main sources of uncertainty. 
In comparison, in sea and lake water $\lambda^{eff} >100-200$ m. 

The Italian NEMO project \cite{NEMO} for a km$^3$ $\nu$ telescope 
has started in 1998 an R\&D activity on the selection of
the optimal site through more than 20 sea campaigns, on
electronics and materials suitable
for long-term undersea measurements, on large area photo-sensors.
More recently the realization of an underwater laboratory (Phase 1 test
site) close to Catania connected to shore by an 28 km-long EOC, where
a couple of prototype towers 
%and a network of 3 junction boxes
will be deployed, has been funded. 
Performance studies have produced a modular detector concept made of 
towers $\sim 600-700$~m high at distances $\gtrsim 120$ m.
%Towers are made by a sequence of storeys interlinked by cables and anchored
%at sea bed and kept vertical by buoys.
Configurations with $\sim 5000$ OMs should
achieve at $E_{\mu} > 100$~TeV
effective areas $>1$~km$^2$ and angular resolutions $<0.1^{\circ}$.
The selected optimal site is Capo Passero, 80 km off-shore  
Catania, 
%where a national laboratory of INFN is located, 
3400 m deep. NEMO performed comparative sea campaigns in collaboration
with ANTARES and Baikal \cite{NEMO}. 
From these campaigns values of the absorption length
of 48 and 66 m for ANTARES and Capo Passero sites, respectively, 
at around 450 nm, have been measured
%where the quantum efficiency of PMTs is maximum, 
and values of 15-30 m in Lake Baikal, while typical values in ice are 
$\gtrsim 100$ m. 
%(strongly dependent on the season).  

Concerning environmental backgrounds,
in sea water continuous rates of the order of tens of kHz are
observed due to $^{40}K$ decay and bursts 
up to several tens of MHz due to living organisms. In ice these backgrounds are
not present, while small contributions at kHz level are due to OM materials.
%Nevertheless the media is less uniform with depth compared to water, light 
%scattering is larger than in water,
%and also the drilling process of strings in the ice produces air bubbles
%around OMs which need to be taken into account.
The longest term measurement ($\sim 100$ d) of optical background was 
performed by ANTARES using the already mentioned 
prototype string. Counting rates show large and 
short lived peaks due to bioluminescence, over a continuous baseline rate
of $\sim 60$ kHz due to $^{40}K$ and bacteria, that varies up to
250 kHz. Studies on correlations between bioluminescence, 
sea currents and string movements are underway. 
At Capo Passero, the NEMO collaboration measured an average rate of 28.5 kHz
in 4 days at a threshold of 0.35 photoelectrons, while the
same device measured 58 kHz at ANTARES site in 4 days. 
Using the 12 PMT floor, NESTOR found that bioluminescence contributes
to the triggered event sample by $\sim 1\%$ of the experiment active time
and with 4-fold coincidence requirements the trigger rate is 2.6 Hz 
(30 mV threshold).
Sedimentation and fouling of optical surfaces causing a 
glass transparency reduction of OM surfaces ($<2\%$ in 1 yr and saturates)
have been measured by ANTARES between July-Dec. 97: the sediment
flux varies between 100-350 mg m$^{-2}$ d$^{-1}$, respectively in
summer and autumn due to clays dragged by rivers. The measured flux
at Capo Passero is 20 mg m$^{-2}$ d$^{-1}$ on average in 40 d, but
longer term measurements are underway.  

\section{Other Techniques}
\label{sec:other}
Electro-magnetic cascades induced by $\nu_{e,\,\tau}$ in dense media
produce coherent radio pulses of Cherenkov radiation
of few ns duration (Askaryan effect \cite{Askarian}) with power
concentrated around Cherenkov angle. 
The RICE detector \cite{RICE} exploits this technique using 16 radio-receivers 
at 100-300 m in AMANDA holes with an effective bandpass of 200-500 MHz
and an attenuation length in ice of $>1$ km. Such a cheap technique
has produced limits competitive to AMANDA ones, even though
further investigations on backgrounds is ongoing. 
This technique is used also for
%also in IceCube holes, possibly at larger depths, 
cascades induced in the lunar regolith and
in salt domes.
Another technique exploits acoustic detection of bipolar pulses 
of $\sim 10$ $\mu$'s duration caused by expansion of
homogeneous media due to energy deposited by showers that converts into heat. 
At typical frequencies of 10 kHz, sound waves
propagate kms in water. Limiting factors are the high directionality
of the acoustic pattern, that restricts the solid angle accessible to sensors,
and the noise due to mammals, wind, thermal noise and human factors.
Currently an acoustic military array close to Bahamas is studying these 
aspects \cite{AUTEC}. Most of the Cherenkov NTs 
also plan to deploy hydrophones. For instance NEMO is going to deploy 4
hydrophones at the test site. 
Finally, methods suitable for $\nu_{\tau}$ detection with fluorescence and
Cherenkov arrays have been proposed,
such as detection of showers induced by $\tau$ leptons
emerging from mountains or Earth-skimming upward $\tau$'s 
decaying after crossing $\sim 10$ kms along Earth cords \cite{Auger}.

These techniques have larger energy thresholds ($\sim 10^{18}$ eV)
than Cherenkov NTs, hence cannot take profit of the atmospheric
$\nu$ measurement for 'calibration' purposes. 
\section{Conclusions}
The current status of NTs has been summarized.
No positive signal has been observed up to now.
The construction of km$^3$ arrays with the proper discovery potentials
is a challenge that is going to start very soon.
\section*{Acknowledgments}
I would like to thank the organizers of the conference, 
particularly W.C. Haxton, and F. Halzen, S. Cecchini and F. Arneodo.
\vskip -0.3 cm

\end{document}